\documentclass[reprint,aps,prd,floats,floatfix,amsmath,amssymb,amsfonts,nofootinbib,longbibliography,superscriptaddress,showkeys]{revtex4-2}

\usepackage{graphicx}
\usepackage{dcolumn}
\usepackage{bm}
\usepackage{natbib}
\usepackage{float}
\usepackage[hyperindex,colorlinks]{hyperref}
\usepackage[capitalise]{cleveref}
\usepackage{orcidlink}
\newcommand{\D}{\text{d}}
\newcommand{\WI}[1]{\textcolor{blue}{#1}}

\DeclareFontFamily{U}{mathb}{\hyphenchar\font45}
\DeclareFontShape{U}{mathb}{m}{n}{
      <5> <6> <7> <8> <9> <10> gen * mathb
      <10.95> mathb10 <12> <14.4> <17.28> <20.74> <24.88> mathb12
      }{}

\newcommand{\eq}{\text{eq}}

\begin{document}

\title{Bulk viscosity from early-time thermalization of cosmic fluids in light of DESI DR2 data}

\author{Hermano Velten\,\orcidlink{0000-0002-5155-7998}}%
\email{hermano.velten@ufop.edu.br}
\affiliation{%
Departamento de F\'isica, Universidade Federal de Ouro Preto (UFOP), Campus Universit\'ario Morro do Cruzeiro, 35.402-136, Ouro Preto, Brazil}%
\affiliation{%
PPGCosmo, Universidade Federal do Esp\'irito Santo, 29075-910, Vit\'oria, ES, Brazil}%

\author{William Iania\,\orcidlink{0009-0004-1966-9211}}
\email{william.iania@edu.ufes.br}
\affiliation{%
PPGCosmo, Universidade Federal do Esp\'irito Santo, 29075-910, Vit\'oria, ES, Brazil}%

\date{\today}


\begin{abstract}
If nonrelativistic dark matter and radiation are allowed to interact, reaching an approximate thermal equilibrium, this interaction induces a bulk viscous pressure changing the effective one-fluid description of the universe dynamics, permitted by the existence of a common temperature. It has been shown that by modelling such components as perfect fluids, a cosmologically relevant bulk viscous pressure, expressed in terms of the Eckart formalism, emerges for dark matter particle masses in the range of $1\,\text{eV} - 10\,\text{eV}$ keeping thermal equilibrium with the radiation. Such a transient bulk viscosity introduces significant effects in the expansion rate near the matter-radiation equality redshift ($z_{\eq} \sim 3400$), impacting also late times leading to a higher inferred value of the Hubble constant $H_0$. Since this mechanism also impacts the sound speed of the baryon-photon fluid, we use the recent DESI DR2 BAO measurements, reported relative to a fiducial $\Lambda$CDM cosmology, to place an upper bound on the logarithm of the free parameter of the model $\tau_\eq$ which represents the time scale in which each component follows its own internal perfect fluid dynamics until thermalization occurs. Our main result is encoded in the bound $\log_{10}(\tau_\eq\,[\rm{s}]) \lesssim -9.76$ (2$\sigma$), with the corresponding dimensionless bulk coefficient $\tilde{\xi}  H_0/H_\eq\lesssim5.94\times10^{-4}$ (2$\sigma$). The obtained constraints show that DESI DR2 data do not support such an interaction between radiation and dark matter prior to the recombination epoch, precluding the model from solving the cosmic tensions.
\end{abstract}

\maketitle

\section{Introduction}

According to the standard cosmological model, in the early universe, radiation dominated the background expansion, consequently suppressing the growth of sub-horizon density fluctuations. As matter energy density took over around the redshift $z_{\eq} \sim 3400$ - the matter-radiation equality epoch - expansion slowed, allowing structure formation to proceed. Radiation at that time consisted of a baryon-photon plasma, where photons redistributed energy through diffusion.
While the cosmological implications of interactions between radiation and dark matter before recombination have been minimally explored, Ref.~\cite{Velten_2021} applied the mechanism introduced by Ref.~\cite{Zimdahl:1996fj} to a model where two adiabatic fluids (matter and radiation) reach approximate thermal equilibrium. In this approach, the thermalized fluids consist of particles whose interactions are sufficiently weak that their interaction energy can be neglected. As a result, the total energy density of the resultant composite fluid can be approximated as the sum of the energy densities of its individual components. Nevertheless, these interactions are sufficiently strong to sustain the equilibrium.

Hence, a nonvanishing bulk viscosity may be produced in the system as a whole, owing to differing cooling rates between the two perfect fluids. This model will be revisited in more detail in the following section. Although interactions are weak enough to neglect their energy contribution, they still produce small nonequilibrium effects, which manifest as an effective bulk viscosity described in terms of Eckart theory, cf. Ref.~\cite{Eckart:1940te}.

Among the various dissipative processes considered in cosmology, bulk viscosity is the most prominently favored. In contrast to shear viscosity and thermal conductivity, it is consistent with the symmetry requirements of homogeneous and isotropic Friedmann–Lemaître–Robertson–Walker (FLRW) spacetimes.
This application highlights the intrinsic nonadiabatic nature of multifluid systems - an aspect absent in the standard cosmological model but widely studied in viscous cosmology scenarios, cf. Refs.~\cite{Zimdahl_1996,Brevik:2005bj,Brevik:2011mm,Li:2009mf,Hipolito-Ricaldi:2010wrq,Velten:2013pra,Brevik:2017msy,PhysRevD.111.083540}. 

The resulting feature is represented by a transient bulk viscous pressure emerging during the radiation-to-matter transition. It fades in both early (ultra-relativistic) and late time (nonrelativistic) limits in agreement with the general theory of nonequilibrium thermodynamics that gives rise to the concept of bulk viscosity, cf. Ref.~\cite{HISCOCK1983466,Weinberg:1972kfs}. However, this feature leaves remarkable imprints on the background expansion rate. Around $z_\eq$ the bulk viscous pressure provides an extra push up into the background dynamics. This effect is designed in Ref.~\cite{Velten_2021} as a transient phenomenon, which vanishes before the decoupling epoch $z_\ast\sim 1100$. Then, there is no damage to the well established Cosmic Microwave Background (CMB) physics, as it was shown in Ref.~\cite{Velten_2021}. In addition, the bulk viscous pressure induces an extra contribution, of nonadiabatic nature, to the effective speed of sound $c_s(z)$ in the baryon-photon fluid. 
 
In order to constrain such an effect, one needs available observational data which are sensitive to early time physics. In this task, Baryon Acoustic Oscillations (BAO) provide a powerful standard ruler for measuring the expansion history of the Universe at different redshifts, cf. Ref.~\cite{Weinberg_2013}. Recent DESI DR2 BAO measurements in Ref.~\cite{DESI:2025zgx} provide cosmic distance ratios in the redshift range from \( z = 0.295 \) to \( z = 2.33 \). The feature imprinted at the drag epoch, denoted by $r_\text{d}$, depends on the sound horizon at that redshift. The computation of $r_\text{d}$ involves the integral of $c_s(z)/H(z)$ from the drag epoch redshift $z_\text{d}$ to the early RD epoch, i.e. $r_\text{d}\equiv \int^{\infty}_{z_\text{d}} \D z(c_s(z)/H(z))$; in the DESI analysis, distances and BAO observables are reported relative to a fiducial $\Lambda$CDM cosmology, adopted as a reference model. Then, the BAO technique has the optimal shape to capture the changes provided by the model we are interested in.

The work is organized as follows.~\Cref{sec:bulk_viscous_cosmology} reviews the formalism of cosmological bulk viscous dynamics. We also calculate in detail the speed of sound $c_s(z)$ of the baryon-photon fluid corrected by the bulk viscous contribution. In~\Cref{sec:statistical_analysis} we describe the cosmological observables we used and give the result of the statistical analysis in form of an upper bound on the model parameter. Quantities evaluated at radiation-matter equality are denoted by the subscript ``eq", observables that refer to the $\Lambda$CDM model are denoted with the subscript ``$\Lambda$''. We focus on flat-background expansion rate, that is, we set the curvature $\Omega_k=0$.

\section{Effective (one-fluid) viscous dynamics from two coupled perfect fluids}
\label{sec:bulk_viscous_cosmology}

\subsection{Matter-radiation thermalization and transient bulk viscosity}
\label{subsec:rm_transient_bulk}

We outline the main steps of the model presented for the first time in Ref.~\cite{Zimdahl:1996fj}, applied to a mixture of radiation and CDM, cf. Ref.~\cite{Velten_2021}. The content of the universe is described by two perfect fluids, identified as radiation ($r$) and nonrelativistic dark matter ($m$). The equilibrium between both components is assumed at a certain time $\eta_0$, then $
T(\eta_0) = T_r(\eta_0) = T_m(\eta_0)$ and $p(\eta_0) = p_r(\eta_0) + p_m(\eta_0)$.
During a subsequent time interval $\tau$, both fluids follow their own internal perfect fluid dynamics. This means that at a time $\eta_0 + \tau$, up to first order
\begin{equation}
\rho_{A}(\eta_0 + \tau) = \rho_{A}(\eta_0) + \tau \dot{\rho}_{A} + \dots\,
\end{equation}
is valid and the subscript $A=(r,m)$ stands for both radiation and nonrelativistic matter. Beyond this point, each fluid evolves according to its own equation of state, and each temperature evolves according to
\begin{equation}\label{eq:temperature_evolution}
\dot{T}_A(\eta_0) = -3H T_A \frac{\partial p_A / \partial T_A}{\partial \rho_A / \partial T_A}\,.
\end{equation}
Then, due to the distinct nature of both fluids, at the instant $\eta_0 + \tau$ their temperature should not be the same $T(\eta_0 + \tau) \neq T_1(\eta_0 + \tau) \neq T_2(\eta_0 + \tau)$, and the sum of the partial pressures reads
\begin{equation}
\label{eq:pdiff}
\begin{split}
    p_r(n_r, T_r) &+ p_m(n_m, T_m) = p_r(n_r, T) + p_m(n_m, T) \\
    &+ (T_r - T) \frac{\partial p_r}{\partial T} + (T_m - T) \frac{\partial p_m}{\partial T}\,,
\end{split}
\end{equation}
being $n$ the particle number density of each fluid. 

For dissipative relativistic fluids described in Eckart's frame, cf. Ref.~\cite{Eckart:1940te}, the energy-momentum tensor at the background is corrected by a nonadiabatic viscous pressure term $\Pi$ such that
\begin{equation}
    T^{\mu\nu}=\rho u^\mu u^\nu + (p+\Pi) h^{\mu\nu}\,,
\end{equation}
with kinetic pressure and energy density related by $p=w\rho$, being $h^{\mu\nu}=u^\mu u^\nu+g^{\mu\nu}$ the projector tensor. Therefore, the difference of the partial pressures stated in~\Cref{eq:pdiff} can be mapped onto a new pressure contribution. It is worth noting that the choice of the latter is in principle free. E.g. a polytropic form could also be proposed. Since we are seeking a phenomenological description of the mechanism, we adopt for simplicity a bulk viscous pressure $\Pi$ expressed within the Eckart formalism; thus
\begin{equation}
    \Pi=
p_r(n_r, T_r) + p_m(n_m, T_m) - p(n, T)
\end{equation} 
results in the corresponding bulk viscous coefficient
\begin{equation}
    \xi = -\tau T \frac{\partial \rho}{\partial T} 
\left( \frac{\partial p_r}{\partial \rho_r} - \frac{\partial p}{\partial \rho} \right) 
\left( \frac{\partial p_m}{\partial \rho_m} - \frac{\partial p}{\partial \rho} \right)\,,
\end{equation}
that within Eckart's framework reads as $\Pi=-\xi \Theta$, where $\xi>0$ is the bulk viscous coefficient and $\Theta=u^{\mu}_{\,\,; \mu}$ is the expansion scalar. Positive values of $\xi$ are required by the second law of thermodynamics, cf. Ref.~\cite{Weinberg:1972kfs}. Consequently, for a FLRW metric $\Pi=-3H\xi$. It encodes the amount of pressure that is opposed to a given volumetric strain rate, whose dimension is the inverse of a time.

The above quantity vanishes in both the ultra-relativistic limit ($p\rightarrow p_r$) and in the nonrelativistic one ($p\rightarrow p_m$).
However, this is not the case around $z_\eq$. The entire cosmic medium can therefore be described by an effective dissipative fluid: different cooling rates give rise to a global term of bulk viscous pressure at the background level. 

Let the radiation have a pressure $p_r = n_r k_\text{B} T_r $ and density $\rho_r = 3 n_r k_\text{B} T_r$, whereas, for the cold dark matter, we change the subscript to $\chi$ avoiding confusion with the mass term, i.e. $p_\chi = n_\chi k_\text{B} T_\chi $ and $\rho_\chi = n_\chi m_\chi c^2 + \frac{3}{2} n_\chi k_\text{B} T_\chi$. Here, $k_\text{B}$ is the Boltzmann constant, $c$ is the speed of light and $m_{\chi}$ the mass of the scalar dark matter particle. One finds (cf. Ref.~\cite{Velten_2021})
\begin{equation}
    \label{eq:xi}
    \xi = \frac{\rho_r}{9}\tilde{\eta}\tau(y)\,, \quad \tilde{\eta}\equiv \frac{\eta_{\chi r}}{2+\eta_{\chi r}}\,,
\end{equation}
with
\begin{equation}\label{eq:eta}
    \eta_{\chi r}\equiv\frac{\eta_{\chi}}{\eta_r}\approx\frac{2.9}{m_{\chi}}[\text{eV}/c^2]\,,
\end{equation}
where $y=a/a_\eq=a(1+z_\eq)$. In the last approximation we have assumed that the mass of a baryon is equal to that of a proton and that the dark matter is nonrelativistic, $m_\chi c^2\gg k_\text{B} T_\chi $. The quantity $\tilde{\eta}$ encodes the relative abundance of dark matter particles $\chi$ in comparison to the photon number. 

The term $\tau(y)$ in~\Cref{eq:xi} is the mean free time for the interaction between the 2 fluids, parametrized as (cf. Ref.~\cite{Velten_2021})
\begin{equation}\label{eq:tau}
    \tau(y) = \tau_\eq\frac{H_\eq}{H_\Lambda}f(y)\,,\quad f(y)\equiv\frac{2y^2}{1+y^4}\,,
\end{equation}
where $H_\Lambda$ is the Hubble rate in the $\Lambda$CDM model, which is later used to give the boundary conditions. The new phenomenological parameter is $\tau_\eq \equiv\tau(y=1)$, i.e. the value $\tau$ assumes at $a_\eq$, where~\cref{eq:tau} reaches its maximum. The fluid description of the universe is valid as long as $\tau H \ll 1$. In light of the previous considerations, the requirement $\tau_\eq H_\eq\ll1$ guarantees the applicability of the fluid-dynamics interpretation to the matter-radiation equality. Thus, the application of the fluid formalism leads to the bound $\tau_\eq\ll10^{-7}\,\mathrm{s}$ (cf. Ref.~\cite{Velten_2021}). This is the first theoretical upper bound on the model parameter. As we shall see later in this work, this is still about two orders of magnitude weaker than the bounds placed by the current observational data. Finally, $f (y)$ is an ad hoc function, necessary to provide a transient behavior, centered around $a_\eq$. The coupling between dark matter and radiation considered in this work does not depend on a specific microscopic interaction model, which would require the introduction of some additional physical parameters beyond the dark matter particle mass. Instead, the interactions only become significant when the energy densities of both fluids are comparable — a feature that naturally arises from the thermodynamical framework adopted by Ref.~\cite{Zimdahl:1996fj}. At earlier (RD) and later (MD) times, when one component dominates the energy budget, the effective interaction vanishes. This behavior is well captured by the parametrization of $f(y)$ and can be visualized in Figure~2 of Ref.~\cite{Velten_2021}. In principle, other prescriptions that constrain $\xi$ to act around the equality could be explored, since its choice is arbitrary and does not depend on the interactions between dark matter and the baryon-photon fluid. We remain agnostic about the model of the microscopic dynamics: it is not relevant in our effective thermodynamical description. It is instead relevant noting that the interaction time - much smaller than the mean free time between the ``collisions'', in turn much smaller than the macroscopic timescale - must be understood as the characteristic time required for interactions to thermalize the fluids and to drive them toward a common equilibrium temperature. This interfluid interaction time is defined within a classical kinetic framework; in light of this, we shall hereafter refer to the interaction time as thermalization time. If the relaxation time is null, the thermalization time is also null, leading to an instantaneous energy transfer, that is unphysical. Hence, $\tau_\eq\to 0$ switches off the interactions between the fluids in order to preserve causality. Consequently, if data constrain the model parameter to be compatible with the null value, such a thermalization could not have happened. This inference comes directly from our assumptions and remains independent of the specific model of the underlying microphysics.

The background expansion dynamics is similar to the standard $\Lambda$CDM model where the total density is written as the sum $\rho=\rho_{m}+\rho_r+\rho_{\Lambda}$. However, there is an additional pressure contribution $\Pi$, such that $p=p_m+p_r+p_{\Lambda}+\Pi$. We consider $p_m=0$, $p_r=\rho_r/3$, and $p_{\Lambda}=-\rho_{\Lambda}$. Such a ``dissipative'' term comes from our particular description, without implying an energy loss from the dark matter-radiation mixture to other components of the total energy-momentum tensor. The energy flux is an internal exchange between dark matter and radiation. The formalism arises when they interact, because the thermalization allows to describe the two-fluid mixture as a unique fluid with a common temperature. Then, during the relaxation time $\tau$, the temperatures of the fluids freely evolve according to~\Cref{eq:temperature_evolution}, giving rise to the partial pressures of~\Cref{eq:pdiff}. In turn, the one-fluid description maps these partial pressure derivatives onto a global, small term of ``dissipation'', that is, a way to express the different evolutions as a single one. If the thermalization does not occur, the two fluids do not reach a thermal equilibrium, there is no unique temperature, and we cannot adopt the one-fluid description.

Therefore, in the presence of a bulk pressure, the continuity equation is given by
\begin{equation}
    \dot{\rho} + 3 H \left( \rho + p_r + p_\Lambda + \Pi \right) = 0\,,
\end{equation}
which, rearranged for the dimensionless expansion rate $E=H/H_0$, gives
\begin{equation}\label{eq:ode_E}
    2Ea\left(  \frac{dE}{da} \right) + 3E^2\left( 1-\frac{\tilde{\xi}}{3E} \right)+\frac{\Omega_{r0}}{a^4}-3\Omega_\Lambda=0\,\WI{.}
\end{equation}
To evaluate the impact of the thermalization (i.e., the bulk viscous coefficient) on the background expansion, we solve~\Cref{eq:ode_E} for \( \tilde{\xi} > 0\), with $E(a_i)=H_\Lambda(a_i)/H_{\Lambda,0}$ as the initial condition, where $a_i = a_\eq/100000$. We adopt \( H_{\Lambda,0}=H_0^{\mathrm{cmb}} = 67.4\, \mathrm{km}\, \mathrm{s}^{-1}\, \mathrm{Mpc}^{-1} \), \(\Omega_{r0}\simeq9.236 \times 10^{-5}\), \(\Omega_{m0}\simeq0.3126\), cf. Ref.~\cite{Planck:2018vyg} and $\Omega_{\Lambda}=1-\Omega_{r0}-\Omega_{m0}$. The $\Lambda$CDM cosmology initializes the evolution of the expansion rate deep in the radiation-dominated era, where the effects on the observables are absent.

The effective dimensionless quantity contributing to the background dynamics is (cf. Ref.~\cite{Velten_2021})
\begin{equation}\label{eq:tilde_xi}
    \frac{3\Pi}{\rho} = -24\pi G H^{-1}\xi\equiv-\frac{\tilde{\xi}}{E}\,,
\end{equation}
where $G$ is the Newton constant. Inserting~\Cref{eq:xi,eq:tau} into~\Cref{eq:tilde_xi} one obtains
\begin{equation}\label{eq:xidimensionless}
    \tilde{\xi} = \tau(y)H_0\Omega_r\tilde{\eta}\,.
\end{equation}
Recall that for a $m_{\chi}=1\,\text{eV}$ dark matter particle our estimation provides $\tilde{\eta}=0.59$. Thus, $\tau_\eq$ is the only free model parameter. The dimensionless bulk viscous coefficient, as given in expression~\Cref{eq:xidimensionless}, features the factor $\tilde{\eta}$, which depends solely on the mass $m_\chi$ of the dark matter particle. In the limit of large mass, $m_\chi \to \infty$, both $\eta_{\chi r}$ and $\tilde{\eta}$ tend to zero, leading to $\xi \to 0$. Thus, this effect does not manifest in extremely massive dark matter candidates, such as WIMPs (Weakly Interacting Massive Particles), whose masses typically lie in the GeV range.

Conversely, for light dark matter candidates with $m_\chi \ll 2.9~\text{eV}/c^2$, the ratio $\eta_{\chi r}$ can become significantly large, causing $\tilde{\eta}$ to approach its maximum value, namely $\tilde{\eta} \to 1$.  At the same time, in order for the estimation $\eta_{\chi r}$ to remain applicable, the nonrelativistic approximation employed for the matter fluid must be valid near the epoch of matter-radiation equality. This condition establishes a lower limit on the mass of dark matter particles: $m_\chi \gtrsim 1~\text{eV}$. We therefore fix $m_\chi = 1\,\rm{eV}$ in the remaining of our analysis, without any loss of generality: we show in~\cref{subsec:dm_particle}, for various fixed $\tau_\eq$, that $m_\chi\sim10\,\rm{eV}$, the heaviest allowed mass, suppresses the corresponding effective bulk viscosity of less than an order of magnitude; such a very small variation guarantees robustness to the choice of fixing it throughout all the analysis.

\begin{figure}[t!]
    \centering
    \includegraphics[width=1\columnwidth]{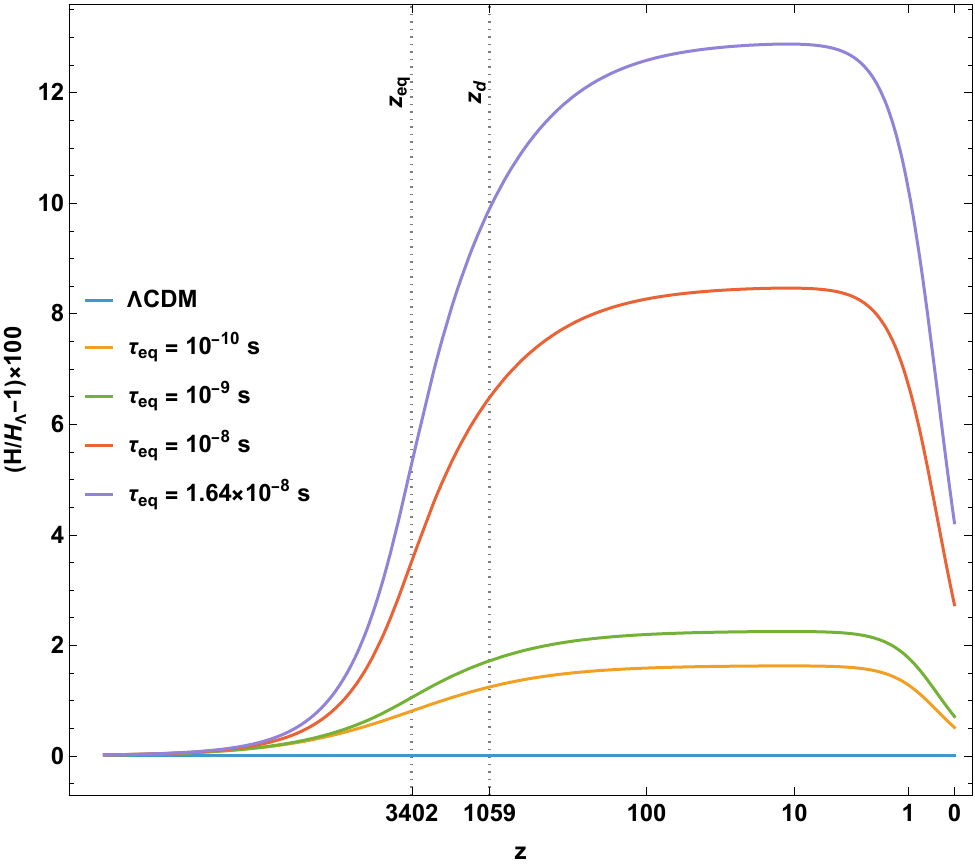}
    \caption{Percentage difference between the Hubble rate computed with and without a bulk viscous pressure, shown as a function of the redshift for different $\tau_\eq$ values. The benchmark $\Lambda$CDM curve corresponds to the viscous case for $\tau_\eq=0\,\text{s}$. The value $\tau_\eq = 1.64 \times 10^{-8}\,\text{s}$ denotes the theoretical upper bound imposed by the positivity of the sound speed, as will be discussed in~\Cref{subsec:sound_speed}.}
    \label{fig:diff_hubble_rates}
\end{figure}

As a result of the numerical evaluation of \Cref{eq:ode_E}, we plot the percentage difference between $H$ and $H_\Lambda$ in~\Cref{fig:diff_hubble_rates} for different values of $\tau_\eq$, showing that for $\tau_\eq\sim 10^{-10}\,\text{s}$, corresponding to $\tilde{\xi}/E\sim10^{-4}$, the $\Lambda$CDM background dynamics is going to be recovered as expected. However, higher values of the parameter lead to a higher $H_0$ compared to $H_0^{\mathrm{cmb}}$. As shown in Ref.~\cite{Velten_2021}, the interval $ 10^{-8}\,\text{s}\lesssim \tau_{\eq}\lesssim 6 \times 10^{-8}\,\text{s}$ can alleviate the Hubble tension by inducing a percentage increase with respect to $H_0^\text{cmb}$.
%

\subsection{Nonadiabatic sound speed of the effective baryon-photon fluid}\label{subsec:sound_speed}

The speed of an oscillation propagating in the baryon-photon fluid is described by the variation of the pressure with respect to the energy density under adiabatic conditions. In the standard cosmological scenario, the following expression works

\begin{equation}\label{eq:sound_speed_definition}
    c_{s,\Lambda}^2=c^2\cdot\left(\frac{\delta p}{\delta\rho}\right)_{\delta S = 0}=\frac{c^2}{3}\frac{1}{1+R(z)}\,,
\end{equation}
with
\begin{equation}\label{eq:relative_momentum_density}
    \quad R(z)=\frac{3}{4}\frac{\rho_{b,0}}{\rho_{\gamma,0}}\frac{1}{1+z}\,,
\end{equation}
where $\rho_{b,0}$ and $\rho_{\gamma,0}$ are the densities of baryons and photons today, respectively, and $\delta S = 0$ ensures that no entropy variation occurs.

In the presence of a bulk viscosity, the kinetic pressure is reduced by the viscous term $\Pi$. Then,~\Cref{eq:sound_speed_definition} is in turn modified by a nonadiabatic correction
\begin{equation}\label{eq:fullcs2}
    c_s^2 = c_{s,\Lambda}^2+c^2\cdot\delta_\rho\Pi\,,
\end{equation}
where $\delta_\rho\Pi \equiv \frac{\delta \Pi}{\delta \rho}<0$. One expects therefore the bulk viscous contribution to reduce the effective value of the squared speed of sound.

Let us start writing the total perturbation for the bulk viscous pressure perturbation as
\begin{equation}
    \delta_\rho\Pi = -3\left[\xi\,\delta_\rho H + H\,\delta_\rho\xi\right]
    = -3\,\delta_\rho\xi\left[\delta_\xi H + H\right]\,.
\end{equation}
We can use for any function $F$ of a generic dynamical variable proportional to the scale factor $x$
\begin{equation}
    \delta F \simeq\frac{\D F}{\D x}\delta x\,,
\end{equation}
thus we compute the following perturbed quantities as intermediate step

\begin{align}
    \delta\rho_r&\simeq-\frac{4\rho_r}{x}\delta x\,,\quad \delta\rho_m\simeq-\frac{3\rho_m}{x}\delta x\,, \\
    \delta\rho&\simeq-\frac{4\rho_r+3\rho_m}{x}\delta x\,,\\
    \delta H_\Lambda&\simeq-\frac{H_0^2}{2xH_\Lambda}\,(4\Omega_{r0}+3\Omega_{m0})\delta x\label{eq:deltaHLambda}\,.
\end{align}
Then, the perturbation of the bulk viscous coefficient with respect to the total energy density fluctuation reads
\begin{equation}
    \begin{split}
        \frac{\delta\xi}{\delta\rho}&\simeq-\,\frac{K\,\rho_r\,x}{(4\rho_r+3\rho_m)\,H_\Lambda}\\
        &\times\left[\frac{\D f}{\D x}-\frac{4f}{x}+ f\,\frac{H_0^2}{2x\,H_\Lambda^{2}}\,(4\Omega_{r0}+3\Omega_{m0})\right]\,, 
    \end{split}
\end{equation}
where
\begin{align}
    \xi&=\dfrac{K}{H_\Lambda}f\rho_r\,,\\
    K&=\frac{\tau_\eq H_\eq\tilde{\eta}}{9}\,,\\
    \frac{\D f}{\D x}&=\frac{4x}{x_{\rm eq}^2}\,\frac{1-(x/x_{\rm eq})^4}{\left[1+(x/x_{\rm eq})^4\right]^2}\,.
\end{align}
The evaluation of $\delta_\xi H$ is more complex than in the $\Lambda$CDM case presented in~\Cref{eq:deltaHLambda}, since the presence of an effective bulk pressure precludes an analytic expression for the Hubble rate. To proceed, it is convenient to express the perturbation of $H$ with respect to the scale factor variable $x$, in order to use~\Cref{eq:ode_E}. We therefore define
\begin{equation}
    A(x)\equiv \frac{\D f}{\D x}-\frac{4f}{x}+ f\,\frac{H_0^2}{2x\,H_\Lambda^{2}}\,(4\Omega_{r0}+3\Omega_{m0})\,,
\end{equation}
so that
\begin{equation}
    \delta\xi\simeq\frac{K}{H_\Lambda}\,\rho_r\,A(x)\,\delta x\,,
\end{equation}
and
\begin{equation}
    \frac{\delta H}{\delta\xi}=\frac{\delta x}{\delta\xi}\frac{\delta H}{\delta x}\simeq\frac{H_\Lambda}{K\,\rho_r\,A(x)}\,\frac{\D H}{\D x}\,,
\end{equation}
with
\begin{equation}
    \begin{split}
        \frac{\D H}{\D x}&=H_0\frac{\D E}{\D x}\\
        &= H_0\left[-\frac{3E}{x}\left(1-\frac{\tilde{\xi}}{3E}\right)+\frac{\Omega_{r0}}{2Ex^5}+\frac{3\Omega_\Lambda}{2Ex}\right].
    \end{split}
\end{equation}
Finally, the complete nonadiabatic correction can be written as
\begin{equation}
    \delta_\rho\Pi \simeq \frac{3K\,\Omega_r\,x}{(4\Omega_r+3\Omega_m)\,H_\Lambda^{2}}
    \left[H\,A(x)+ f(x)\,\frac{\D H}{\D x}\right]\,,
\end{equation}
where the densities are expressed in their dimensionless form.

\begin{figure}[t!]
    \centering
    \includegraphics[width=1\columnwidth]{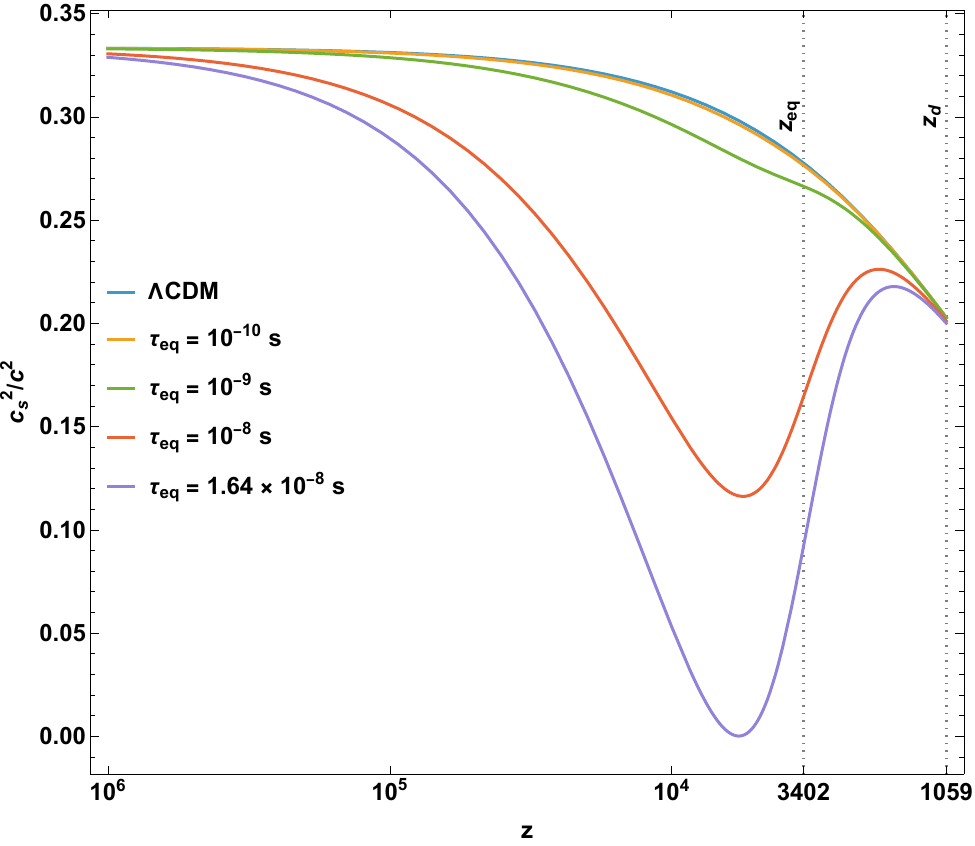}
    \caption{Evolution of the squared sound speed \Cref{eq:fullcs2}, including the nonadiabatic correction, as a function of the redshift for different $\tau_\eq$ values. In the limit $z\to\infty$, the radiation-dominated behavior $c_s^2=1/3$ is recovered, while at $z=z_\text{d}$, baryons decouple from the photon drag and the acoustic oscillation freeze out, leaving the characteristic BAO imprint. For $\tau_\eq= 1.64 \times10^{-8}\,\text{s}$, the adiabatic contribution is almost completely suppressed by the nonadiabatic component leading to a vanishing $c^2_s$ before $z_\eq$. For $\tau_\eq\sim10^{-10}\,\text{s}$, the nonadiabatic correction is almost negligible.}
    \label{fig:diff_sound_speeds}
\end{figure}

The effective squared speed of sound in~\Cref{eq:fullcs2} should therefore take into account the above expression for the $\delta_{\rho}\Pi$ contribution. Since the condition $c_s^2>0$ must hold at all times, we determine the maximum value that ensures it across all redshifts. According to this criterion, the corresponding upper bound on $\tau_\eq$ is found to be
\begin{equation}
    \tau_\eq<1.64\times10^{-8}\,\text{s}\,,
\end{equation}
yielding the corresponding upper bound on the bulk dimensionless parameter
\begin{equation}\label{eq:taueq_csbound}
    \max_{\text{all }z}{\frac{\tilde{\xi}}{E}}=\frac{\tilde{\xi}}{E}\Bigg\vert_{z_\eq}<0.052\,.
\end{equation}
In~\Cref{fig:diff_sound_speeds} we plot the squared speed of sound (in $c$ units) as a function of the redshift. It is clear that the bulk viscous contribution reduces the squared speed of sound of the total effective cosmological fluid before the equality epoch when the transient effect starts. At later times the effect fades away.

\section{Cosmological constraints on the thermalization time scale at matter-radiation equality}
\label{sec:statistical_analysis}

Our primary goal in this section is to use BAO observables and constraints available in the recent DESI DR2 dataset to place an upper bound on the free model parameter $\tau_\eq$. We compare the results with DES Y3 data of Ref.~\cite{DES:2021esc}.

\subsection{The role of the mass of the dark matter particles}
\label{subsec:dm_particle}

%
\begin{figure}[t!]
    \centering
    \includegraphics[width=0.96\columnwidth]{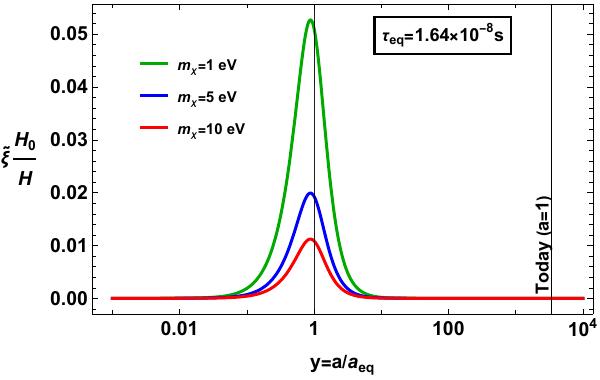}
    \includegraphics[width=\columnwidth]{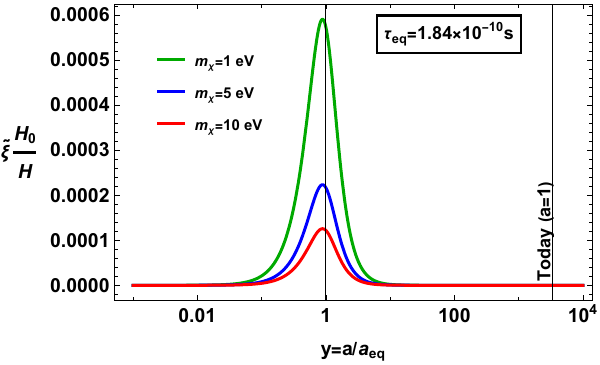}
    \caption{Effective bulk viscosity as a function of the normalized scale factor, for different various of $m_\chi$ in the allowed range, at fixed $\tau_\eq$. The upper panel shows the results for the maximum value of $\tau_\eq$, constrained by the positivity of the speed of sound in~\Cref{eq:taueq_csbound}. In the lower one we use the $2\sigma$ upper bound given by DESI DR2 data in~\Cref{eq:finalbound}.}
    \label{fig:varying_mchi}
\end{figure}
Although the model of Ref.~\cite{Velten_2021} allows the dark matter particles to have a mass in the range $1-10\,\rm{eV}$, where either the fluid-dynamical description and the nonrelativistic limit are valid, we fix $m_\chi=1\,\rm{eV}$ in the analysis. This assumption is reasonable in light of the analytical form of the (dimensionless) bulk viscosity and its dependence on such a mass. Using~\Cref{eq:xi,eq:eta,eq:tau,eq:xidimensionless} we have $\tilde{\xi} = \tau(y)H_0\Omega_r\tilde{\eta}\sim\tau_\eq \cdot\tilde{\eta}(m_\chi)$, with $0.13\lesssim\tilde{\eta}(m_\chi)\lesssim0.59$ in the considered mass range; lower $m_\chi$ correspond to higher $\tilde{\eta}$. Then, even choosing the heaviest allowed mass, the effective bulk viscosity $\tilde{\xi}/E$-that directly modifies the expansion rate evolution in~\Cref{eq:ode_E}-is correspondingly suppressed by a factor $\sim0.13$, less than an order of magnitude, without altering significantly any statistical conclusion on the upper bounds on $\tau_\eq$. Then, we further confirm our choice to fix the mass, and sample only on $\log_{10}(\tau_\eq)$, that, differently from $m_\chi$, can span over several orders of magnitude, highly featuring the viscous coefficient and the subsequent effect on the BAO observables.~\Cref{fig:varying_mchi} shows the behavior of the bulk viscous coefficient for various values of $m_\chi$, taking the highest possible value of $\tau_\eq$ and the $2\sigma$ upper bound from DESI, confirming the solidity of our choice for the value of the dark matter particle mass.

\subsection{BAO distances}
\label{subsec:bulk_rd}

Distributions of galaxies show an enhancement at a certain characteristic scale, determined by the physics of the sound waves in the pre-recombination universe; gravity pulled matter into dense regions, while the pressure from photons pushed it away, creating sound waves that rippled through the plasma. These baryon ripples, or Baryon Acoustic Oscillations (BAO), froze at the end of baryon drag epoch, giving rise to the typical scale of clustering enhancement, the \textit{sound horizon at drag epoch}
\begin{equation}\label{eq:rd}
    r_\text{d} = \int_{z_\text{d}}^{\infty}\D z \frac{c_s(z)}{H(z)}=\int_0^{a_\text{d}}\D a \frac{c_s(a)}{a^2 H(a)}\,.
\end{equation}
The end of the drag epoch, at $z_\text{d} \approx 1060$, cf. Ref.~\cite{Planck:2018vyg}, marks the redshift at which baryons cease to be dynamically coupled to the photon fluid: photons no longer \textit{drag} them, so baryons are free to fall into gravitational potentials. The distinction with recombination is important. Between $z_\ast$ and $z_\text{d}$, photons are already nearly free-streaming, yet baryons still feel residual radiation pressure. From a heuristic point of view, one may think of baryons as still propagating in an effective ``photon medium'' that drags them until $z_\text{d}$.

The sound horizon is used as the cosmic ladder and other cosmic distances can be defined in units of $r_\text{d}$. The distances we use henceforth are the Hubble distance
\begin{equation}\label{eq:DH}
    D_H(z)=\frac{c}{H(z)}\,,
\end{equation}
the comoving transverse distance for a flat universe
\begin{equation}\label{eq:DM}
    \begin{split}
        D_M(z)&=\frac{c}{H_0\sqrt{\Omega_K}}\sinh{\left[\sqrt{\Omega_K}\int_0^z\frac{\D z^\prime}{E(z^\prime)} \right]}\Bigg\vert_{|\Omega_K|\ll1} \\
        &=\frac{c}{H_0}\int_0^z\frac{\D z^\prime}{E(z^\prime)}\quad\text{(flat universe)}\,,
    \end{split}
\end{equation}
and the isotropic volume-averaged distance
\begin{equation}\label{eq:DV}
    D_V(z)=\left[zD_M(z)^2D_H(z)\right]^{1/3}\,.
\end{equation}
The key BAO distance combinations provided by the DESI collaboration, reported relative to a fiducial $\Lambda$CDM cosmology, are $D_V(z)/r_{\rm d}$ for isotropic analyses and $D_M(z)/r_{\rm d}$ and $D_H(z)/r_{\rm d}$ for anisotropic measurements; from these quantities, the derived ratio $D_M(z)/D_H(z)$ can also be constructed.

\subsection{Bayesian analysis}
\label{subsec:bayesian_analysi}

%
\begin{figure}[t!]
    \centering
    \includegraphics[width=1\columnwidth]{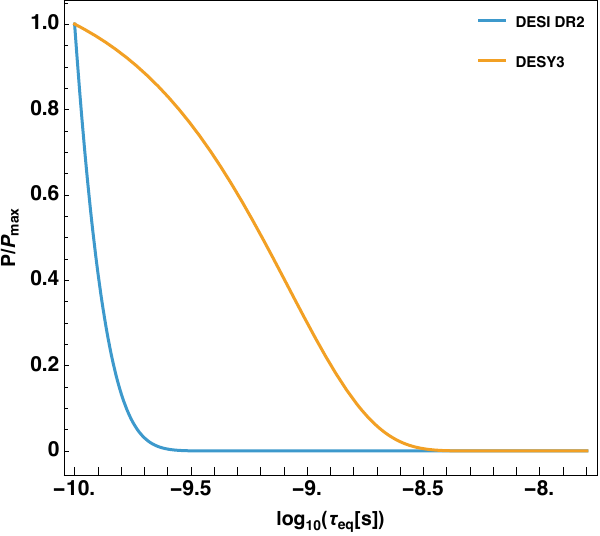}
    \caption{1D marginalized posteriors, normalized to their maximum value, for the combination of redshift bins of DESI DR2, in comparison with DES Y3 measurement of the comoving transverse distance.}
    \label{fig:ppmax}
\end{figure}
We perform a Bayesian analysis on the logarithm of the parameter $\log_{10}(\tau_\eq)$, replacing $\tau_\eq=10^{\log_{10}(\tau_\eq)}$; we use the observables in Table~IV of Ref.~\cite{DESI:2025zgx}, denoted by $\mathbf{d}_\text{obs}$. The corresponding observables computed in the model are $\mathbf{d}_\text{th}(\log_{10}(\tau_\eq))$. For each redshift bin, they are defined as
\begin{equation}
    \mathbf{d}_\text{obs}\equiv\left[\frac{D_V}{r_\text{d}},\frac{D_M}{D_H},\frac{D_M}{r_\text{d}},\frac{D_H}{r_\text{d}}\right]\,,
\end{equation}
and
\begin{equation}
    \begin{split}
        \mathbf{d}_\text{th}(\log_{10}(\tau_\eq))&\equiv
\Bigg[\frac{D_V(\log_{10}(\tau_\eq))}{r_\text{d}(\log_{10}(\tau_\eq))},\frac{D_M(\log_{10}(\tau_\eq))}{D_H(\log_{10}(\tau_\eq))},\\
&\frac{D_M(\log_{10}(\tau_\eq))}{r_\text{d}(\log_{10}(\tau_\eq))},\frac{D_H(\log_{10}(\tau_\eq))}{r_\text{d}(\log_{10}(\tau_\eq))}\Bigg]\,.
    \end{split}
\end{equation}
We build the likelihood from the $\chi^2$ statistics
\begin{equation}
    \chi^2 = \left(\mathbf{d}_\text{obs}-\mathbf{d}_\text{th}\right)^\text{T}\cdot\mathbf{C}^{-1}\cdot\left(\mathbf{d}_\text{obs}-\mathbf{d}_\text{th}\right)\,,
\end{equation}
where $\textbf{C}^{-1}$ denotes the inverse of the BAO covariance matrix. Cross-bin correlations are neglected, as in the main DESI analysis, and errors are considered Gaussian.

We impose a uniform prior, and give as the only physical information its maximum, determined by the logarithm of the upper bound in~\Cref{sec:bulk_viscous_cosmology}; the minimum can be considered equivalent to $\Lambda$CDM in light of the results shown in~\Cref{fig:diff_hubble_rates,fig:diff_sound_speeds}
\begin{equation}
    \pi\left(\log_{10}(\tau_\eq\,[\rm{s}])\right)\sim\mathcal{U}\left(-10,-7.785156\right)\,.
\end{equation}
The one-dimensional posteriors normalized to their maximum are shown in~\Cref{fig:ppmax}. The posteriors are plotted for each tracer (or redshift bin), and for the joint DESI DR2 dataset, with a further comparison with the DES Y3 measurement of the comoving transverse distance at an effective redshift of $z_{eff}=0.835$, given by $D_M/r_\text{d}=18.92\pm0.51$ in Ref.~\cite{DES:2021esc}.

\begin{figure}[t!]
    \centering
    \includegraphics[width=1\columnwidth]{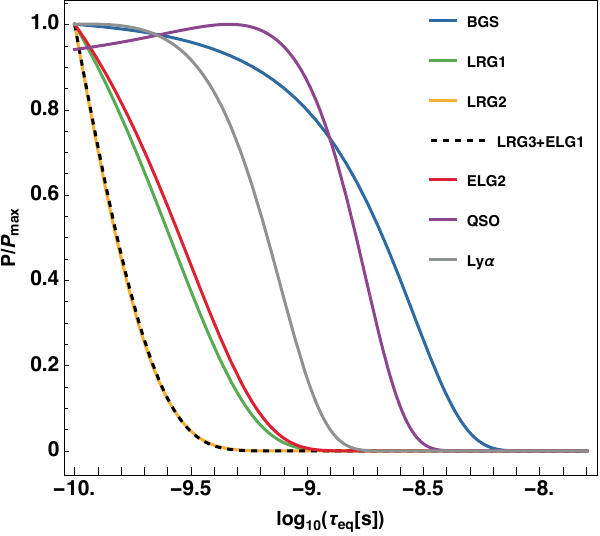}
    \caption{1D marginalized posteriors, normalized to their maximum value, for the single tracers of DESI DR2 dataset.}
    \label{fig:ppmax_alltracers}
\end{figure}

\Cref{tab:bin_sigma} shows the results for each single redshift bin, for their combination, and for the DES Y3 angular distance, through the $1\sigma-2\sigma$ upper bounds on the logarithm of the relaxation time, demonstrating consistency across the redshift range and with the DES distance ratio; in~\Cref{fig:ppmax_alltracers} we can see that all the tracers yield a posterior peaking on the lowest value, except for the QSO curve, nevertheless, its 2$\sigma$ bound is lower than the bound of BGS. Their joint analysis reads
\begin{equation}\label{eq:finalbound}
    \log_{10}(\tau_\eq\,[\rm{s}])\lesssim-9.76\quad \text{($2 \sigma$) -  DESI DR2}\,,
\end{equation}
which implies an upper bound on the effective dimensionless bulk viscous coefficient that modifies the background dynamics
\begin{equation}\label{eq:finalboundxi}
\max_{\text{all }z}{\frac{\tilde{\xi}}{E}}=\frac{\tilde{\xi}}{E}\Bigg\vert_{z_\eq}\lesssim 5.94\times10^{-4}\quad \text{($2 \sigma$) - DESI DR2}\,.
\end{equation}
%

\section{Conclusions}
\label{sec:conclusions}
%
\begin{table}[t!]
    \centering
    \begin{ruledtabular}
    \begin{tabular}{l c c c}
        Tracer or dataset     & $z_{\rm eff}$ & $\log_{10}(\tau_{\rm eq}\,[\rm{s}])\lesssim1\sigma$ & $\log_{10}(\tau_{\rm eq}\,[\rm{s}])\lesssim2\sigma$ \\
        BGS         & 0.295  & -9.10  & -8.58 \\
        LRG1        & 0.510  & -9.64 & -9.33 \\
        LRG2        & 0.706  & -9.80 & -9.58 \\
        LRG3+ELG1   & 0.934  & -9.80 & -9.58 \\
        ELG2        & 1.321  & -9.62 & -9.30 \\
        QSO         & 1.484  & -9.18 & -8.77  \\
        Ly$\alpha$  & 2.330  & -9.43 & -9.09  \\
        \textbf{DESI DR2} & \textbf{All} & -9.90 & \textbf{-9.76} \\
        DES Y3      & 0.835  & -9.39 & -8.93  \\
    \end{tabular}
    \end{ruledtabular}
    \caption{Marginalized constraints for each redshift bin of DESI DR2~\cite{DESI:2025zgx} and DES Y3~\cite{DES:2021esc}, expressed as upper bounds on $\tau_\eq$ at the 68\% ($1\sigma$) and 95\% ($2\sigma$) credible regions.}
    \label{tab:bin_sigma}
\end{table}
We have revisited the mechanism proposed in Ref.~\cite{Zimdahl:1996fj}, which establishes a novel expression for cosmological bulk viscosity arising from thermal equilibrium between two adiabatic fluids. This framework was later applied in Ref.~\cite{Velten_2021} to the early universe, where the cosmic substratum is modelled as a combination of pressureless dark matter and a radiation fluid. The resulting formulation yields an effective bulk viscosity for a one-fluid description of the cosmic energy budget. That is, although composed of two adiabatic components, the global expansion acquires an additional non-adiabatic feature.
In light of recent BAO data from the second release of the DESI survey, we are now able to impose tighter constraints on the magnitude of this thermalization process occurring prior to the recombination epoch. To derive these bounds, we introduced a new expression for the effective speed of sound in the baryon–photon fluid under the influence of this mechanism. As discussed in~\Cref{subsec:sound_speed}, this proves to be a highly sensitive probe. Our statistical analysis yields an upper bound - cf.~\Cref{eq:finalbound} - on the magnitude of the transient bulk viscosity around the matter–radiation equality epoch. In terms of the bulk viscous coefficient, this is expressed in~\Cref{eq:finalboundxi}. The allowed effect is small enough to yield negligible contributions to the overall cosmological expansion. Based on the perturbative analysis presented in Ref.~\cite{Velten_2021}, which focuses on scalar cosmological perturbations within the same model, we conclude that any interaction between dark matter and radiation of this type prior to recombination can be ruled out. Consequently, while in Ref.~\cite{Velten_2021} it is shown that $\tau_\eq\simeq7\cdot10^{-8}\,\rm{s}$ could lead to a strong alleviation of the cosmic tensions, providing a higher value of $H_0$ and also a lower value of the matter clustering parameter $S_8\equiv\sigma_8\sqrt{\Omega_{m0}/0.3}$, for $\tau_\eq\lesssim 10^{-9}\,\rm{s}$ the mitigation effect is negligible, (cf. their Figures 3-4). Thereby, the $2\sigma$ upper bound from DESI of $\sim10^{-10}\,\rm{s}$ cancels out the possibility to solve the cosmic tensions through this model of thermalization: this is the main result of our work.

Finally, it is important to note that the model examined here is based on the Eckart formalism, which is known to suffer from further issues of non-causality and instability. Although the physical motivation that leads us to claim that DESI data do not support the model is independent of the particular form of the bulk viscous pressure, the specific adopted formalism is valid at the background level and gives rise to the standard problems of Eckart theory of relativistic dissipative fluids. Future investigations into bulk viscous effects arising from interactions among cosmic components may benefit from adopting the fully causal M\"uller–Israel–Stewart (MIS) theory, cf. Refs.~\cite{1967ZPhy..198..329M,ISRAEL1976310,Israel:1976efz}, or the intermediate first-order relativistic hydrodynamics framework developed in Refs.~\cite{Disconzi_2015,Bemfica:2017wps,Bemfica:2019cop}, which is locally well-posed - i.e. causal, stable, and strongly hyperbolic - similar to the MIS theory.

Furthermore, the analysis was restricted to a mixture of radiation and CDM: the general outline of the approach is expected to be valid for other possible interactions between energy components. Another possible application could be a thermalization process between dark matter and dark energy, if the latter is represented in the fluid picture. We leave this analysis for a future work.

\acknowledgments{The authors thank FAPEMIG, FAPES, CAPES and CNPq for financial support. The authors would like to acknowledge the use of the computational resources provided by the \href{https://computacaocientifica.ufes.br/scicom}{Sci-Com Lab} of the Department of Physics at UFES, which was funded by FAPES, CAPES and CNPq. The authors acknowledge Winfried Zimdahl for insightful discussions.}

%



\begin{thebibliography}{22}%
\makeatletter
\providecommand \@ifxundefined [1]{%
 \@ifx{#1\undefined}
}%
\providecommand \@ifnum [1]{%
 \ifnum #1\expandafter \@firstoftwo
 \else \expandafter \@secondoftwo
 \fi
}%
\providecommand \@ifx [1]{%
 \ifx #1\expandafter \@firstoftwo
 \else \expandafter \@secondoftwo
 \fi
}%
\providecommand \natexlab [1]{#1}%
\providecommand \enquote  [1]{``#1''}%
\providecommand \bibnamefont  [1]{#1}%
\providecommand \bibfnamefont [1]{#1}%
\providecommand \citenamefont [1]{#1}%
\providecommand \href@noop [0]{\@secondoftwo}%
\providecommand \href [0]{\begingroup \@sanitize@url \@href}%
\providecommand \@href[1]{\@@startlink{#1}\@@href}%
\providecommand \@@href[1]{\endgroup#1\@@endlink}%
\providecommand \@sanitize@url [0]{\catcode `\\12\catcode `\$12\catcode `\&12\catcode `\#12\catcode `\^12\catcode `\_12\catcode `\%12\relax}%
\providecommand \@@startlink[1]{}%
\providecommand \@@endlink[0]{}%
\providecommand \url  [0]{\begingroup\@sanitize@url \@url }%
\providecommand \@url [1]{\endgroup\@href {#1}{\urlprefix }}%
\providecommand \urlprefix  [0]{URL }%
\providecommand \Eprint [0]{\href }%
\providecommand \doibase [0]{https://doi.org/}%
\providecommand \selectlanguage [0]{\@gobble}%
\providecommand \bibinfo  [0]{\@secondoftwo}%
\providecommand \bibfield  [0]{\@secondoftwo}%
\providecommand \translation [1]{[#1]}%
\providecommand \BibitemOpen [0]{}%
\providecommand \bibitemStop [0]{}%
\providecommand \bibitemNoStop [0]{.\EOS\space}%
\providecommand \EOS [0]{\spacefactor3000\relax}%
\providecommand \BibitemShut  [1]{\csname bibitem#1\endcsname}%
\let\auto@bib@innerbib\@empty
\bibitem [{\citenamefont {Velten}\ \emph {et~al.}(2021)\citenamefont {Velten}, \citenamefont {Costa},\ and\ \citenamefont {Zimdahl}}]{Velten_2021}%
  \BibitemOpen
  \bibfield  {author} {\bibinfo {author} {\bibfnamefont {H.}~\bibnamefont {Velten}}, \bibinfo {author} {\bibfnamefont {I.}~\bibnamefont {Costa}},\ and\ \bibinfo {author} {\bibfnamefont {W.}~\bibnamefont {Zimdahl}},\ }\bibfield  {journal} {\bibinfo  {journal} {Physical Review D}\ }\textbf {\bibinfo {volume} {104}},\ \href {https://doi.org/10.1103/physrevd.104.063507} {10.1103/physrevd.104.063507} (\bibinfo {year} {2021})\BibitemShut {NoStop}%
\bibitem [{\citenamefont {Zimdahl}(1996{\natexlab{a}})}]{Zimdahl:1996fj}%
  \BibitemOpen
  \bibfield  {author} {\bibinfo {author} {\bibfnamefont {W.}~\bibnamefont {Zimdahl}},\ }\href {https://doi.org/10.1093/mnras/280.4.1239} {\bibfield  {journal} {\bibinfo  {journal} {Mon. Not. Roy. Astron. Soc.}\ }\textbf {\bibinfo {volume} {280}},\ \bibinfo {pages} {1239} (\bibinfo {year} {1996}{\natexlab{a}})},\ \Eprint {https://arxiv.org/abs/astro-ph/9602128} {arXiv:astro-ph/9602128} \BibitemShut {NoStop}%
\bibitem [{\citenamefont {Eckart}(1940)}]{Eckart:1940te}%
  \BibitemOpen
  \bibfield  {author} {\bibinfo {author} {\bibfnamefont {C.}~\bibnamefont {Eckart}},\ }\href {https://doi.org/10.1103/PhysRev.58.919} {\bibfield  {journal} {\bibinfo  {journal} {Phys. Rev.}\ }\textbf {\bibinfo {volume} {58}},\ \bibinfo {pages} {919} (\bibinfo {year} {1940})}\BibitemShut {NoStop}%
\bibitem [{\citenamefont {Zimdahl}(1996{\natexlab{b}})}]{Zimdahl_1996}%
  \BibitemOpen
  \bibfield  {author} {\bibinfo {author} {\bibfnamefont {W.}~\bibnamefont {Zimdahl}},\ }\href {https://doi.org/10.1103/physrevd.53.5483} {\bibfield  {journal} {\bibinfo  {journal} {Physical Review D}\ }\textbf {\bibinfo {volume} {53}},\ \bibinfo {pages} {5483–5493} (\bibinfo {year} {1996}{\natexlab{b}})}\BibitemShut {NoStop}%
\bibitem [{\citenamefont {Brevik}\ and\ \citenamefont {Gorbunova}(2005)}]{Brevik:2005bj}%
  \BibitemOpen
  \bibfield  {author} {\bibinfo {author} {\bibfnamefont {I.~H.}\ \bibnamefont {Brevik}}\ and\ \bibinfo {author} {\bibfnamefont {O.}~\bibnamefont {Gorbunova}},\ }\href {https://doi.org/10.1007/s10714-005-0178-9} {\bibfield  {journal} {\bibinfo  {journal} {Gen. Rel. Grav.}\ }\textbf {\bibinfo {volume} {37}},\ \bibinfo {pages} {2039} (\bibinfo {year} {2005})},\ \Eprint {https://arxiv.org/abs/gr-qc/0504001} {arXiv:gr-qc/0504001} \BibitemShut {NoStop}%
\bibitem [{\citenamefont {Brevik}\ \emph {et~al.}(2011)\citenamefont {Brevik}, \citenamefont {Elizalde}, \citenamefont {Nojiri},\ and\ \citenamefont {Odintsov}}]{Brevik:2011mm}%
  \BibitemOpen
  \bibfield  {author} {\bibinfo {author} {\bibfnamefont {I.}~\bibnamefont {Brevik}}, \bibinfo {author} {\bibfnamefont {E.}~\bibnamefont {Elizalde}}, \bibinfo {author} {\bibfnamefont {S.}~\bibnamefont {Nojiri}},\ and\ \bibinfo {author} {\bibfnamefont {S.~D.}\ \bibnamefont {Odintsov}},\ }\href {https://doi.org/10.1103/PhysRevD.84.103508} {\bibfield  {journal} {\bibinfo  {journal} {Phys. Rev. D}\ }\textbf {\bibinfo {volume} {84}},\ \bibinfo {pages} {103508} (\bibinfo {year} {2011})},\ \Eprint {https://arxiv.org/abs/1107.4642} {arXiv:1107.4642 [hep-th]} \BibitemShut {NoStop}%
\bibitem [{\citenamefont {Li}\ and\ \citenamefont {Barrow}(2009)}]{Li:2009mf}%
  \BibitemOpen
  \bibfield  {author} {\bibinfo {author} {\bibfnamefont {B.}~\bibnamefont {Li}}\ and\ \bibinfo {author} {\bibfnamefont {J.~D.}\ \bibnamefont {Barrow}},\ }\href {https://doi.org/10.1103/PhysRevD.79.103521} {\bibfield  {journal} {\bibinfo  {journal} {Phys. Rev. D}\ }\textbf {\bibinfo {volume} {79}},\ \bibinfo {pages} {103521} (\bibinfo {year} {2009})},\ \Eprint {https://arxiv.org/abs/0902.3163} {arXiv:0902.3163 [gr-qc]} \BibitemShut {NoStop}%
\bibitem [{\citenamefont {Hipolito-Ricaldi}\ \emph {et~al.}(2010)\citenamefont {Hipolito-Ricaldi}, \citenamefont {Velten},\ and\ \citenamefont {Zimdahl}}]{Hipolito-Ricaldi:2010wrq}%
  \BibitemOpen
  \bibfield  {author} {\bibinfo {author} {\bibfnamefont {W.~S.}\ \bibnamefont {Hipolito-Ricaldi}}, \bibinfo {author} {\bibfnamefont {H.~E.~S.}\ \bibnamefont {Velten}},\ and\ \bibinfo {author} {\bibfnamefont {W.}~\bibnamefont {Zimdahl}},\ }\href {https://doi.org/10.1103/PhysRevD.82.063507} {\bibfield  {journal} {\bibinfo  {journal} {Phys. Rev. D}\ }\textbf {\bibinfo {volume} {82}},\ \bibinfo {pages} {063507} (\bibinfo {year} {2010})},\ \Eprint {https://arxiv.org/abs/1007.0675} {arXiv:1007.0675 [astro-ph.CO]} \BibitemShut {NoStop}%
\bibitem [{\citenamefont {Velten}\ \emph {et~al.}(2013)\citenamefont {Velten}, \citenamefont {Schwarz}, \citenamefont {Fabris},\ and\ \citenamefont {Zimdahl}}]{Velten:2013pra}%
  \BibitemOpen
  \bibfield  {author} {\bibinfo {author} {\bibfnamefont {H.}~\bibnamefont {Velten}}, \bibinfo {author} {\bibfnamefont {D.~J.}\ \bibnamefont {Schwarz}}, \bibinfo {author} {\bibfnamefont {J.~C.}\ \bibnamefont {Fabris}},\ and\ \bibinfo {author} {\bibfnamefont {W.}~\bibnamefont {Zimdahl}},\ }\href {https://doi.org/10.1103/PhysRevD.88.103522} {\bibfield  {journal} {\bibinfo  {journal} {Phys. Rev. D}\ }\textbf {\bibinfo {volume} {88}},\ \bibinfo {pages} {103522} (\bibinfo {year} {2013})},\ \Eprint {https://arxiv.org/abs/1307.6536} {arXiv:1307.6536 [astro-ph.CO]} \BibitemShut {NoStop}%
\bibitem [{\citenamefont {Brevik}\ \emph {et~al.}(2017)\citenamefont {Brevik}, \citenamefont {Gr{\o}n}, \citenamefont {de~Haro}, \citenamefont {Odintsov},\ and\ \citenamefont {Saridakis}}]{Brevik:2017msy}%
  \BibitemOpen
  \bibfield  {author} {\bibinfo {author} {\bibfnamefont {I.}~\bibnamefont {Brevik}}, \bibinfo {author} {\bibfnamefont {{\O}.}~\bibnamefont {Gr{\o}n}}, \bibinfo {author} {\bibfnamefont {J.}~\bibnamefont {de~Haro}}, \bibinfo {author} {\bibfnamefont {S.~D.}\ \bibnamefont {Odintsov}},\ and\ \bibinfo {author} {\bibfnamefont {E.~N.}\ \bibnamefont {Saridakis}},\ }\href {https://doi.org/10.1142/S0218271817300245} {\bibfield  {journal} {\bibinfo  {journal} {Int. J. Mod. Phys. D}\ }\textbf {\bibinfo {volume} {26}},\ \bibinfo {pages} {1730024} (\bibinfo {year} {2017})},\ \Eprint {https://arxiv.org/abs/1706.02543} {arXiv:1706.02543 [gr-qc]} \BibitemShut {NoStop}%
\bibitem [{\citenamefont {Paul}(2025)}]{PhysRevD.111.083540}%
  \BibitemOpen
  \bibfield  {author} {\bibinfo {author} {\bibfnamefont {T.}~\bibnamefont {Paul}},\ }
  \bibfield  {title} {\enquote {\bibinfo {title} {Origin of bulk viscosity in cosmology and its thermodynamic implications},}\ }
  \bibfield  {journal} {\bibinfo  {journal} {Phys. Rev. D}\ }\textbf {\bibinfo {volume} {111}},\ \bibinfo {pages} {083540} (\bibinfo {year} {2025}),\
  \href {https://doi.org/10.1103/PhysRevD.111.083540} {10.1103/PhysRevD.111.083540}\BibitemShut {NoStop}%
\bibitem [{\citenamefont {Hiscock}\ and\ \citenamefont {Lindblom}(1983)}]{HISCOCK1983466}%
  \BibitemOpen
  \bibfield  {author} {\bibinfo {author} {\bibfnamefont {W.~A.}\ \bibnamefont {Hiscock}}\ and\ \bibinfo {author} {\bibfnamefont {L.}~\bibnamefont {Lindblom}},\ }\href {https://doi.org/https://doi.org/10.1016/0003-4916(83)90288-9} {\bibfield  {journal} {\bibinfo  {journal} {Annals of Physics}\ }\textbf {\bibinfo {volume} {151}},\ \bibinfo {pages} {466} (\bibinfo {year} {1983})}\BibitemShut {NoStop}%
\bibitem [{\citenamefont {Weinberg}(1972)}]{Weinberg:1972kfs}%
  \BibitemOpen
  \bibfield  {author} {\bibinfo {author} {\bibfnamefont {S.}~\bibnamefont {Weinberg}},\ }\href@noop {} {\emph {\bibinfo {title} {{Gravitation and Cosmology}: {Principles and Applications of the General Theory of Relativity}}}}\ (\bibinfo  {publisher} {John Wiley and Sons},\ \bibinfo {address} {New York},\ \bibinfo {year} {1972})\BibitemShut {NoStop}%
\bibitem [{\citenamefont {Weinberg}\ \emph {et~al.}(2013)\citenamefont {Weinberg}, \citenamefont {Mortonson}, \citenamefont {Eisenstein}, \citenamefont {Hirata}, \citenamefont {Riess},\ and\ \citenamefont {Rozo}}]{Weinberg_2013}%
  \BibitemOpen
  \bibfield  {author} {\bibinfo {author} {\bibfnamefont {D.~H.}\ \bibnamefont {Weinberg}}, \bibinfo {author} {\bibfnamefont {M.~J.}\ \bibnamefont {Mortonson}}, \bibinfo {author} {\bibfnamefont {D.~J.}\ \bibnamefont {Eisenstein}}, \bibinfo {author} {\bibfnamefont {C.}~\bibnamefont {Hirata}}, \bibinfo {author} {\bibfnamefont {A.~G.}\ \bibnamefont {Riess}},\ and\ \bibinfo {author} {\bibfnamefont {E.}~\bibnamefont {Rozo}},\ }\href {https://doi.org/10.1016/j.physrep.2013.05.001} {\bibfield  {journal} {\bibinfo  {journal} {Physics Reports}\ }\textbf {\bibinfo {volume} {530}},\ \bibinfo {pages} {87–255} (\bibinfo {year} {2013})}\BibitemShut {NoStop}%
\bibitem [{\citenamefont {Abdul~Karim}\ \emph {et~al.}(2025)\citenamefont {Abdul~Karim} \emph {et~al.}}]{DESI:2025zgx}%
  \BibitemOpen
  \bibfield  {author} {\bibinfo {author} {\bibfnamefont {M.}~\bibnamefont {Abdul~Karim}} \emph {et~al.} (\bibinfo {collaboration} {DESI}),\ }\href@noop {} {\  (\bibinfo {year} {2025})},\ \Eprint {https://arxiv.org/abs/2503.14738} {arXiv:2503.14738 [astro-ph.CO]} \BibitemShut {NoStop}%
\bibitem [{\citenamefont {Abbott}\ \emph {et~al.}(2022)\citenamefont {Abbott} \emph {et~al.}}]{DES:2021esc}%
  \BibitemOpen
  \bibfield  {author} {\bibinfo {author} {\bibfnamefont {T.~M.~C.}\ \bibnamefont {Abbott}} \emph {et~al.} (\bibinfo {collaboration} {DES}),\ }\href {https://doi.org/10.1103/PhysRevD.105.043512} {\bibfield  {journal} {\bibinfo  {journal} {Phys. Rev. D}\ }\textbf {\bibinfo {volume} {105}},\ \bibinfo {pages} {043512} (\bibinfo {year} {2022})},\ \Eprint {https://arxiv.org/abs/2107.04646} {arXiv:2107.04646 [astro-ph.CO]} \BibitemShut {NoStop}%
\bibitem [{\citenamefont {Aghanim}\ \emph {et~al.}(2020)\citenamefont {Aghanim} \emph {et~al.}}]{Planck:2018vyg}%
  \BibitemOpen
  \bibfield  {author} {\bibinfo {author} {\bibfnamefont {N.}~\bibnamefont {Aghanim}} \emph {et~al.} (\bibinfo {collaboration} {Planck}),\ }\href {https://doi.org/10.1051/0004-6361/201833910} {\bibfield  {journal} {\bibinfo  {journal} {Astron. Astrophys.}\ }\textbf {\bibinfo {volume} {641}},\ \bibinfo {pages} {A6} (\bibinfo {year} {2020})},\ \bibinfo {note} {[Erratum: Astron.Astrophys. 652, C4 (2021)]},\ \Eprint {https://arxiv.org/abs/1807.06209} {arXiv:1807.06209 [astro-ph.CO]} \BibitemShut {NoStop}%
\bibitem [{\citenamefont {{M{\"u}ller}}(1967)}]{1967ZPhy..198..329M}%
  \BibitemOpen
  \bibfield  {author} {\bibinfo {author} {\bibfnamefont {I.}~\bibnamefont {{M{\"u}ller}}},\ }\href {https://doi.org/10.1007/BF01326412} {\bibfield  {journal} {\bibinfo  {journal} {Zeitschrift fur Physik}\ }\textbf {\bibinfo {volume} {198}},\ \bibinfo {pages} {329} (\bibinfo {year} {1967})}\BibitemShut {NoStop}%
\bibitem [{\citenamefont {Israel}(1976)}]{ISRAEL1976310}%
  \BibitemOpen
  \bibfield  {author} {\bibinfo {author} {\bibfnamefont {W.}~\bibnamefont {Israel}},\ }\href {https://doi.org/https://doi.org/10.1016/0003-4916(76)90064-6} {\bibfield  {journal} {\bibinfo  {journal} {Annals of Physics}\ }\textbf {\bibinfo {volume} {100}},\ \bibinfo {pages} {310} (\bibinfo {year} {1976})}\BibitemShut {NoStop}%
\bibitem [{\citenamefont {Israel}\ and\ \citenamefont {Stewart}(1976)}]{Israel:1976efz}%
  \BibitemOpen
  \bibfield  {author} {\bibinfo {author} {\bibfnamefont {W.}~\bibnamefont {Israel}}\ and\ \bibinfo {author} {\bibfnamefont {J.~M.}\ \bibnamefont {Stewart}},\ }\href {https://doi.org/10.1016/0375-9601(76)90075-X} {\bibfield  {journal} {\bibinfo  {journal} {Phys. Lett. A}\ }\textbf {\bibinfo {volume} {58}},\ \bibinfo {pages} {213} (\bibinfo {year} {1976})}\BibitemShut {NoStop}%
\bibitem [{\citenamefont {Disconzi}\ \emph {et~al.}(2015)\citenamefont {Disconzi}, \citenamefont {Kephart},\ and\ \citenamefont {Scherrer}}]{Disconzi_2015}%
  \BibitemOpen
  \bibfield  {author} {\bibinfo {author} {\bibfnamefont {M.~M.}\ \bibnamefont {Disconzi}}, \bibinfo {author} {\bibfnamefont {T.~W.}\ \bibnamefont {Kephart}},\ and\ \bibinfo {author} {\bibfnamefont {R.~J.}\ \bibnamefont {Scherrer}},\ }\bibfield  {journal} {\bibinfo  {journal} {Physical Review D}\ }\textbf {\bibinfo {volume} {91}},\ \href {https://doi.org/10.1103/physrevd.91.043532} {10.1103/physrevd.91.043532} (\bibinfo {year} {2015})\BibitemShut {NoStop}%
\bibitem [{\citenamefont {Bemfica}\ \emph {et~al.}(2018)\citenamefont {Bemfica}, \citenamefont {Disconzi},\ and\ \citenamefont {Noronha}}]{Bemfica:2017wps}%
  \BibitemOpen
  \bibfield  {author} {\bibinfo {author} {\bibfnamefont {F.~S.}\ \bibnamefont {Bemfica}}, \bibinfo {author} {\bibfnamefont {M.~M.}\ \bibnamefont {Disconzi}},\ and\ \bibinfo {author} {\bibfnamefont {J.}~\bibnamefont {Noronha}},\ }\href {https://doi.org/10.1103/PhysRevD.98.104064} {\bibfield  {journal} {\bibinfo  {journal} {Phys. Rev. D}\ }\textbf {\bibinfo {volume} {98}},\ \bibinfo {pages} {104064} (\bibinfo {year} {2018})},\ \Eprint {https://arxiv.org/abs/1708.06255} {arXiv:1708.06255 [gr-qc]} \BibitemShut {NoStop}%
\bibitem [{\citenamefont {Bemfica}\ \emph {et~al.}(2019)\citenamefont {Bemfica}, \citenamefont {Disconzi},\ and\ \citenamefont {Noronha}}]{Bemfica:2019cop}%
  \BibitemOpen
  \bibfield  {author} {\bibinfo {author} {\bibfnamefont {F.~S.}\ \bibnamefont {Bemfica}}, \bibinfo {author} {\bibfnamefont {M.~M.}\ \bibnamefont {Disconzi}},\ and\ \bibinfo {author} {\bibfnamefont {J.}~\bibnamefont {Noronha}},\ }\href {https://doi.org/10.1103/PhysRevLett.122.221602} {\bibfield  {journal} {\bibinfo  {journal} {Phys. Rev. Lett.}\ }\textbf {\bibinfo {volume} {122}},\ \bibinfo {pages} {221602} (\bibinfo {year} {2019})},\ \Eprint {https://arxiv.org/abs/1901.06701} {arXiv:1901.06701 [gr-qc]} \BibitemShut {NoStop}%
\end{thebibliography}
\end{document}